\documentclass[floats,floatfix,showpacs,amssymb,prd,twocolumn,superscriptaddress,nofootinbib,nolongbibliography,reprint]{revtex4-2}

\usepackage{amssymb,amsmath,verbatim,mathtools,needspace,enumitem,etoolbox,graphicx,physics,microtype,afterpage,bigints,gensymb,tabularx,xspace}
\usepackage[usenames,dvipsnames]{xcolor}
\definecolor{linkcolor}{rgb}{0.0,0.3,0.5}
\usepackage[colorlinks = true,
            linkcolor = linkcolor,
            urlcolor  = linkcolor,
            citecolor = linkcolor,
            anchorcolor = linkcolor]{hyperref}   
\usepackage{orcidlink}
\usepackage{aasmacros}
\usepackage[commandnameprefix=ifneeded, authormarkup=none,final]{changes}

\definechangesauthor[name={AU}, color=red]{au} %
\newcommand{\aur}[2]{\replaced[id=au]{#1}{#2}} %
\newcommand{\aud}[1]{\deleted[id=au]{#1}} %
\newcommand{\aua}[1]{\added[id=au]{#1}} %

\newcommand{\ssim}{\mathchar"5218\relax\,}

\begin{document}

\title{
Stars or gas? Constraining the hardening processes \\ of   massive black-hole binaries with LISA
}

\newcommand{\bham}{\affiliation{Institute for Gravitational Wave Astronomy \& School of Physics and Astronomy, University of Birmingham, Birmingham, B15 2TT, UK}}
\newcommand{\milan}{\affiliation{Dipartimento di Fisica ``G. Occhialini'', Universit\'a degli Studi di Milano-Bicocca, Piazza della Scienza 3, 20126 Milano, Italy}}
\newcommand{\infn}{\affiliation{INFN, Sezione di Milano-Bicocca, Piazza della Scienza 3, 20126 Milano, Italy}}  

\author{Alice Spadaro~\orcidlink{0009-0000-8104-6171}}
\email{a.spadaro3@campus.unimib.it} 
\milan \infn
\author{Riccardo Buscicchio~\orcidlink{0000-0002-7387-6754}}
\milan \infn
\bham
\author{David Izquierdo--Villalba~\orcidlink{0000-0002-6143-1491}}
\milan \infn
\author{Davide Gerosa~\orcidlink{0000-0002-0933-3579}}
\milan \infn
\author{Antoine Klein~\orcidlink{0000-0001-5438-9152}}
\bham
\author{Geraint Pratten~\orcidlink{0000-0003-4984-0775}}
\bham

\date{\today}

\begin{abstract}
Massive black-hole binaries will be the loudest sources detectable by LISA. 
These systems are predicted to form during the hierarchical assembly of cosmic structures and coalesce by interacting with the surrounding environment.
The hardening phase of their orbit is driven by either stars or gas and encodes distinctive features into the binary black holes  that can potentially be reconstructed with gravitational-wave observations.
We present a Bayesian framework to assess the likelihood of massive mergers being hardened by either  gaseous or stellar interactions. 
We use state-of-the-art astrophysical models tracking the cosmological evolution of massive black-hole binaries and construct a large number of simulated catalogs of sources detectable by LISA.  
From these, we select a representative catalog and run both parameter estimation assuming a realistic LISA response as well \aua{as} model comparison capturing selection effects.
Our results suggest that, at least within the context of the adopted models, future LISA observations can confidently constrain whether stars or gas are responsible for the binary hardening.
We stress that accurate astrophysical modeling of the black-hole spins and the inclusion of subdominant emission modes in the adopted signal might be crucial to avoid systematic biases.
\end{abstract}

\maketitle

\section{\label{sec:intro} Introduction}
According to the current $\Lambda$CDM model of structure formation, galaxies form hierarchically~\cite{1978MNRAS.183..341W}.
In this context, the formation of massive black-hole (BH) binaries 
is a direct consequences of galaxy mergers~\cite{1980Natur.287..307B}.
Present evidence for massive BH binaries relies on observational signatures of active galactic nuclei, including offset or double-peaked spectral lines and periodic variations in light curves~\cite{2019NewAR..8601525D,2022LRR....25....3B}. 
While electromagnetic observations lack the capability to directly resolve massive BH binaries below the parsec scale, the recent evidence for a nano-Hertz stochastic gravitational-wave (GW) signal in pulsar timing array datasets points to the existence of a large cosmic population of high mass ($>10^{7}\rm{M_{\odot}}$) merging BH binaries~\cite{2023ApJ...951L...8A,2023A&A...678A..50E,2023ApJ...951L...6R,2023RAA....23g5024X}.
The detection of individual mergers of $10^{4}-10^{7}\rm{M_{\odot}}$ BHs is a cornerstone in the science case of future  space-based  GW detectors, notably the Laser Interferometer Space Antenna (LISA)~\cite{2024arXiv240207571C}.

The astrophysical environments surrounding massive BH binaries are crucial in shaping the binary dynamics and promoting the final coalescence~\cite{2023LRR....26....2A}. 
In particular, mergers can only be explained if dissipation mechanisms of astrophysical nature are at play in the early inspiral, before binaries \aur{enter}{enters} their GW-driven regime. 
Modeling such processes is an active field of research, as different assumptions lead to different predictions for both properties and rates of merging massive BHs.
Broadly speaking, hardening mechanisms can be divided into ``gas hardening'' and ``stellar hardening''~\cite{2012AdAst2012E...3D}. 
The former, which dominates in gas-rich environments, is driven by the interaction of the massive BH binary with a gaseous circumbinary disk~\cite{2008ApJ...672...83M,2009MNRAS.393.1423C}.
Conversely, the latter takes place primarily in gas-poor environments, where the binary orbital separation shrinks due to three-body interactions with individual stars~\cite{1996NewA....1...35Q}.
Both gas and stellar hardening affect the merger rate~\cite{2013CQGra..30x4009S,2016PhRvD..93b4003K} and can potentially be reconstructed with LISA observations~\cite{2021PhRvD.104h3027T}. 
\setlength{\parskip}{2pt}

With this in mind, we present a Bayesian framework to compare single detections of massive BH binaries from LISA-simulated catalogs against state-of-the-art astrophysical simulations. 
We use the \textsc{L-Galaxies} semi-analytic model~\cite{2015MNRAS.451.2663H} applied to the high-resolution \textsc{Millenium-II} dark-matter (DM) merger trees.
We specifically investigate whether it will be possible to infer that putative massive BH binaries observed by LISA have evolved in either gas or stellar environments.
We construct realistic LISA catalogs, perform parameter estimation including the full LISA response with the \textsc{Balrog} \cite{2023PhRvD.108l4045P,2023PhRvD.107l3026P,2023PhRvD.108l3029S} code, and implement a Bayesian model comparison between the two hardening mechanisms while considering selection effects~\cite{2023MNRAS.525.3986M}.

This paper is organized as follows. In Sec.~\ref{sec:astro}, we illustrate the adopted astrophysical models. 
In Sec.~\ref{sec:lisacat}, we present the construction of our LISA catalogs. 
In Sec.~\ref{sec:inf}, we describe our parameter-estimation and model-selection strategies.
In Sec.~\ref{sec:results}, we present and discuss our findings. 
Finally, in Sec.~\ref{sec:conclusions} we summarize our results and outline possible future developments.
Throughout this paper, we use units where $c = 1$.

\section{\label{sec:astro} Astrophysical models}

\subsection{\label{sec:lgalaxies} Galaxy evolution model}

We use the \textsc{L-Galaxies} semi-analytical model as presented in Refs.~\cite{2015MNRAS.451.2663H,2022MNRAS.509.3488I,2023MNRAS.518.4672S}. 
In particular, the model  is built on top of merger trees from the \textsc{Millenium-II}~\cite{2009MNRAS.398.1150B}
cosmological simulation, which follows the evolution of $2160^3$ dark matter particles with a mass of $6.885\times10^{6}{\rm M_{\odot}}/h$ within a comoving box of side $100~{\rm Mpc}/h$, later rescaled~\cite{2010MNRAS.405..143A} to match the Planck cosmology~\cite{2014A&A...571A..16P} (here $h = 10^{-2} H_0$ Mpc s/km and $H_0$ is the Hubble constant).

\textsc{L-Galaxies} associates a fraction of baryonic matter to each newly resolved DM halo in the form of a diffuse, spherical, and quasi-static hot gas atmosphere. As the gas cools down, it settles into a disk~\cite{1993ApJS...88..253S}; this facilitates episodes of star formation, which in turn results in the assembly of a stellar disk component. Galaxies are also allowed to assemble an overdensity of stars in their central regions (i.e. their bulges). This is triggered by either internal processes, namely non-axisymmetric instabilities that redistribute the stellar matter, or galaxy mergers.

\subsection{\label{sec:mbhbs} Black-hole growth}

\aur{For each newly resolved DM halo, \textsc{L-Galaxies} assess whether physical conditions are favorable for the formation of a massive BH seed. 
The model incorporates different processes, including direct collapse of pristine gas clouds, runaway stellar mergers, and PopIII star formation~\cite{2023MNRAS.518.4672S}. 
As a result, the initial BH  mass function arises as a combination of different formation channels, with masses ranging from $\ssim10^2$ to $\ssim10^5~{\rm M_\odot}$~\cite{2023MNRAS.518.4672S}.}{Each newly resolved DM halo is initialized with a massive BH seed spanning a mass range from $\ssim10^2$ to $\ssim10^5~{\rm M_\odot}$~\cite{2023MNRAS.518.4672S}.}
The initial dimensionless spin parameter $\chi$ is  set uniformly in $[0, 0.998]$~\cite{1974ApJ...191..507T}.

The evolution of masses and spins is influenced by the accretion of surrounding gas into the BHs as well as BH-binary coalescences. 
In particular, gas accretion is triggered by both galaxy mergers and disk instabilities~\cite{2008ApJ...684..822B,2011MNRAS.410...53F,2012MNRAS.423.2533B}, and predominantly drives the mass growth~\cite{2014MNRAS.440.1590D}.
The cold gas available for accretion settles \aua{around the BH} in a reservoir of mass $M_{\rm{res}}$, \aua{which is proportional to the cold gas mass of the galaxy~\cite{2024A&A...686A.183I}.} \aud{around the BH, which} 
\aua{The gas reservoir} is then progressively consumed through a sequence of transient accretion disks~\cite{2020MNRAS.495.4681I,2022MNRAS.509.3488I}.
As for the BH spin, its value is set by the frequency of accretion events consumed in prograde or retrograde orbits, which in \textsc{L-Galaxies} is linked to the coherence of the bulge kinematics~\cite{2013ApJ...762...68D,2014ApJ...794..104S}.
The contribution of binary coealescences to the BH mass and spin evolution is important at low redshifts ($z<2$), where galaxies have mostly depleted their gas reservoirs through star formation and feedback processes.

The post-merger BH masses and spins are computed using fits to numerical-relativity simulations~\cite{2008PhRvD..78h1501T,2009ApJ...704L..40B}.
During the BH evolution, \textsc{L-Galaxies} tracks the evolution of the spin magnitude $\chi$ but not the spin direction, which is however important for LISA observations.
In this work, we construct this property for each binary component using the gas fraction of the environment $f_{\rm gas}=M_{\rm{res}}/(M_{\rm{res}} + M)$, where $M$ is the total mass of the BH binary. 
If $f_{\rm gas}>0.5$, coalescence occurs in a gas-rich environment and the two spins are assumed to align to the binary orbital angular momentum via the Bardeen-Petterson effect~\cite{1975ApJ...195L..65B,2010MNRAS.402..682D,2020MNRAS.496.3060G}. 
If instead $f_{\rm gas}<0.5$, the environment is gas-poor and the BH spins are assumed to be isotropically distributed~\cite{2007ApJ...661L.147B,2014MNRAS.440.1590D}. 
\begin{figure}
    \centering
    \includegraphics[width=1.\columnwidth, keepaspectratio]{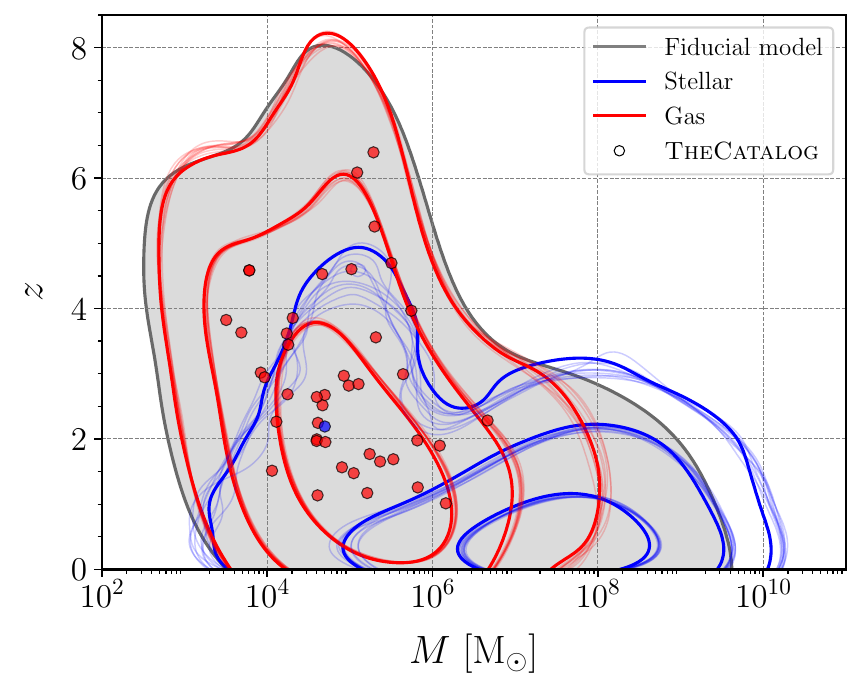}
    \caption{Density distribution of the total mass $M$ and redshift $z$ of massive BH binary mergers. Our fiducial population is shown in gray, where the contour encloses 99\% of the estimated source density. The red (blue) distribution corresponds to the sub-population of mergers hardened by gas (stars); contours contain 50\%, 90\%, and 99\% of the estimated source density.
    Thick contours refer to additional realizations to capture the statistical fluctuations in the underlying astrophysical model.
    Circles denote the 44 sources from \textsc{TheCatalog} used in our analysis, with colors indicating their respective $f_{\rm{gas}}$ values.
    Note the two pairs of overlapping sources, c.f. Fig.~\ref{fig:waterfallplot}.
    }
    \label{fig:popdensity_gas_st}
\end{figure}

\subsection{\label{sec:hardening} Binary hardening}

The dynamical pathway of massive BH binaries can be  divided into three phases: pairing, hardening, and GW inspiral~\cite{1980Natur.287..307B}.
Following a galaxy merger, the pairing phase reduces the BH-BH separation from $\ssim \rm{kpc}$ to $\ssim\rm{pc}$ through dynamical friction.
This is implemented in \textsc{L-Galaxies} as in Ref.~\cite{2008gady.book.....B}.
As the binary forms, different physical processes contribute to the shrinking of the binary semi-major axis depending on the gas fraction $f_{\rm{gas}}$.
For $f_{\rm{gas}}>~\!\!0.5$, gas hardening dominates and the decay of the binary orbit is set by accretion torques from the circumbinary disk. 
In gas-poor environments with $f_{\rm{gas}}<~\!\!0.5$, instead, the binary evolution is driven by capture and ejection of stars. 

Gravitational-radiation reaction becomes dominant at roughly milliparsec scales and drives the system to the merger.
For details on the \textsc{L-Galaxies} implementation see Ref.~\cite{2022MNRAS.509.3488I}.

\subsection{Fiducial population}

The model described in the previous subsections provides the astrophysical population of BH mergers we refer to as ``fiducial'', which is depicted in grey in Fig.~\ref{fig:popdensity_gas_st} and presented in Ref.~\cite{2024A&A...686A.183I}.
\begin{figure}
    \centering
    \includegraphics[width=1.\columnwidth, keepaspectratio]{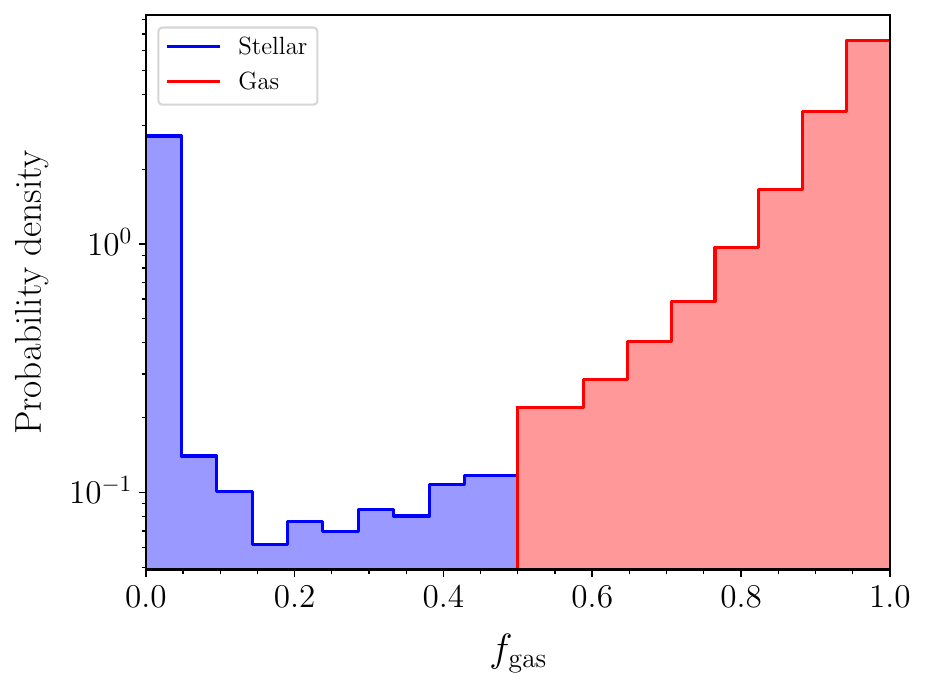}
    \caption{Distribution of the gas fraction  $f_{\rm gas}$ of the environment across the fiducial population. Sources in blue (red) with $f_{\rm gas}<0.5 $ ($f_{\rm gas}>0.5$) are classified as mergers occurring in a gas-poor (gas-rich) environment.
    The two sub-population distributions are jointly normalized to the full population.
}
    \label{fig:hist_fgas}
\end{figure}
This population comprises two distinct sub-populations with peaks at $f_{\rm gas}\ssim0$ (stellar hardening) and $f_{\rm gas}\ssim1$ (gas hardening) parameter (Fig.~\ref{fig:hist_fgas}). 
The latter dominates the overall merger rate by a factor of~$\ssim5$. 
Figure~\ref{fig:popdensity_gas_st} shows the respective mass and redshift distributions for each sub-populations.
Systems evolving in gaseous environments merge on average at higher redshifts $z\lesssim 8$ and have lower masses $M \sim 10^{4}-10^{6}~{\rm M_{\odot}}$ compared to binaries hardened by stellar interactions which instead are more prominent at $ z\lesssim 3$ and have $M\sim 10^{5}-10^{9}~{\rm M_{\odot}}$. 
Nonetheless, the two populations overlap significantly in the $M-z$ parameter space, making their distinguishability with LISA non trivial. 
\begin{figure*}
\centering
\includegraphics[width=\textwidth]{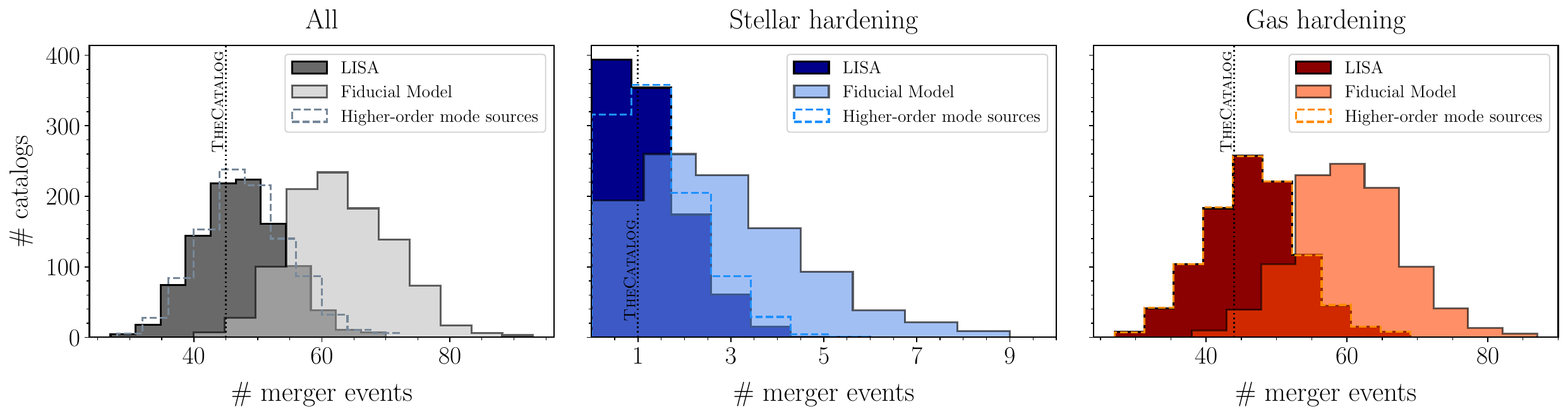}
\caption{Distribution of massive BH binary mergers across the generated catalogs. The left (gray), middle (blue), and right (red) panel shows the full population of BHs, the sub-population of stellar-hardened sources, and the sub-population of gas-hardened sources, respectively. Light histograms show the astrophysical distributions [steps (i)$-$(iii) in Sec.~\ref{sec:lisacat}] while dark histograms show the distributions of merger events detectable by LISA during an observation time $T=5{\rm yr}$ [steps (iv)-(vii) in Sec.~\ref{sec:lisacat}]. The dashed distributions are a variation of the latter where we include higher-order harmonics and remove the condition on $f_{\rm cut}$ [step (iv)]. The vertical dotted line denotes the selected catalog used to perform the statistical analysis of Sec.~\ref{sec:inf}.
}
\label{fig:catalogs_distribution}
\end{figure*}

\section{\label{sec:lisacat} Mock LISA catalogs}

We build mock LISA observation catalogs on top of the \textsc{L-Galaxies} simulations. The latter provides the number density of massive BH binary mergers, $\dd^2 N/ \dd z \dd V_{c}$, per unit redshift $z$ and comoving volume $V_{c}$. 
We proceed as follows:
\begin{enumerate}[label=(\roman*)]
    \item We compute the total merger rate of massive BH mergers predicted by the fiducial model~\cite{2020ApJ...904...16B}
    \begin{equation}
        \frac{\dd N}{\dd t_{\rm obs}}=\int{\frac{\dd N}{\dd z}~w(z)~\dd z},
    \end{equation}
    where 
    \begin{equation}
        w(z)=4\pi\left[\frac{d_{\rm L}}{(1+z)}\right]^{2}\frac{\dd z}{\dd V_{c}}.
        \label{eq:weights}
    \end{equation}
In the equations above, $d_{\rm L}$ is the luminosity distance to the source and $t_{\rm obs}$ is the time measured at the detector.
    \item We draw the number of sources $N_{\rm{astro}}$ according to a Poisson distribution with mean value $\lambda=T\times \dd N/ \dd t_{\rm obs}$ where $T$ is the duration of the observing period. 
    Massive BH binary signals in LISA typically last hours to months and gradually accumulate signal-to-noise ratio (SNR) over their inspiral. 
    We conservatively set $T=5\rm{yr}$ in the calculation above, which accommodates for signals merging after the nominal duration $T_{\rm{LISA}}=4\rm{yr}$ \cite{2024arXiv240207571C}.
    \item We extract $N_{\rm{astro}}$ sources from the astrophysical population with relative weights given by $w_{i}=~\!\!w(z_i)$.  
    Each source is characterized by the binary component masses $m_{1,2}$, 
    \vspace{-0.07em}
    mass ratio $q=m_2/m_1$, redshift $z$, and the aligned dimensionless spins $\chi_{1,2}$ \aua{$=\vec{\chi}_{1,2}\cdot\hat{L}$ where $\vec{\chi}_{1,2}$ are the dimensionless BH spins and $\hat{L}$ is the unit vector along the direction of the orbital angular momentum}.
    \item From these, we further select sources with a cutoff frequency $f_{\rm cut}>1\times10^{-4}~\rm{Hz}$ and $q>~\!\!0.01$. We set $f_{\rm cut}=5f_{\rm ISCO}$ where $f_{\rm ISCO}$ is the detector-frame GW frequency at innermost stable circular orbit.
    This is to ensure that the dominant mode is in the LISA sensitivity band and that the adopted waveform model is  reliable.
    \item We use the  \textsc{IMRPhenomXHM}~\cite{2020PhRvD.102f4002G} waveform approximant which captures the full coalescence of quasi-circular, non-precessing BH binaries.
    The harmonics $h_{lm}$ are calibrated to numerical relativity and include the ($l,|m|)=~\!\!\{(2,2),(2,1),(3,3),(3,2),(4,4)\}$ multipoles.
    The implementation of the LISA response to such GW signal in the \textsc{Balrog} code has been presented in Ref.~\cite{2023PhRvD.107l3026P}.
    \item  For each source, we draw its extrinsic parameters by sampling uniformly the time to merger  $t_m\in [0,5 {\rm yr}]$, the initial phase $\phi_0\in[0,2\pi]$, the polarization angle $\psi\in[-\pi/2,\pi/2]$, the ecliptic sine latitude $\sin\beta\in~\!\![-1,1]$ , the ecliptic longitude $\lambda\in[0,2\pi]$, and the  cosine inclination $\cos\iota\in[-1,1]$.
    \item The SNR of each source, given its parameters $\theta$, is evaluated using the three noise-orthogonal TDI channels $h=\{h_k; k=A,E,T\}$ as follows 
    \begin{equation}
        \mathrm{SNR}^2 = \sum\limits_{k}\big\langle h_k(\theta) \big| h_k(\theta) \big\rangle_k.
        \label{eq:snr}
    \end{equation}
    We assume constant and equal-armlength approximation~\cite{2021LRR....24....1T}. 
    The inner product is given by
    \begin{equation}
        \big<a|b\big>_k=2\int_{f_{\mathrm{min}}}^{f_{\mathrm{max}}}\frac{\tilde{a}(f)\tilde{b}^*(f)+\tilde{a}^*(f)\tilde{b}(f)}{S_k(f)}\dd f, 
         \end{equation}
    where $\tilde{a}(f)$ denotes the Fourier transform of the time series $a(t)$ and $S_k(f)$ the noise power spectral density of the $k$-th TDI channel.
    For the latter, we use the semi-analytical expression of Ref.~\cite{2017PhRvD..95j3012B} which models the superposition of LISA stationary instrumental noise and astrophysical confusion noise from unresolved Galactic binaries. We use a low-frequency cut-off $f_{\rm{min}}=0.1~\rm{mHz}$, as set in the ESA definition study report~\cite{2024arXiv240207571C}, and integrate up to $f_{\rm{max}}=1\rm{Hz}$, which is well above the maximum frequency of all generated GW signals. We consider sources detectable if they exceed the threshold $\rm{SNR}=10$~\cite{2024arXiv240207571C}. 
    \end{enumerate}
We iterate the above procedure 1000 times, resulting in a distribution of LISA catalogs. 
The corresponding distributions for the number of merger events is shown in Fig.~\ref{fig:catalogs_distribution}.
Specifically, steps (i)$-$(iii) generate the light-gray histogram in the left panel, depicting the distribution of astrophysical mergers predicted by the fiducial model. 
The dark-gray histogram is obtained by applying steps (iv)$-$(vii) to the astrophysical catalogs and shows the distribution of massive BH merger events detectable by LISA assuming the dominant $l=2$, $|m|=2$ emission mode. 
The blue (red) histograms in the middle (right) panel show the subpopulations of mergers that evolved via stellar (gas)-hardening.
Overall, the LISA catalogs have about one order of magnitude fewer events compared to the astrophysical catalogs, which is due to the detection criteria described above. 
The observable catalogs predominantly feature massive BH binary mergers occurring in gas-rich environments (46 on average) compared to systems evolving in gas-poor environments (1 on average).
The averaged count of detectable sources mildly increases to 50 when we relax the assumption on the mass ratio, allowing $q$ to be as small as 0.001.
These sources, and those at even lower mass ratio, border the extreme mass-ratio inspiral  regime and will need to be treated differently; we leave this to future work.
\begin{figure*}
    \centering
    \includegraphics[width=\textwidth]{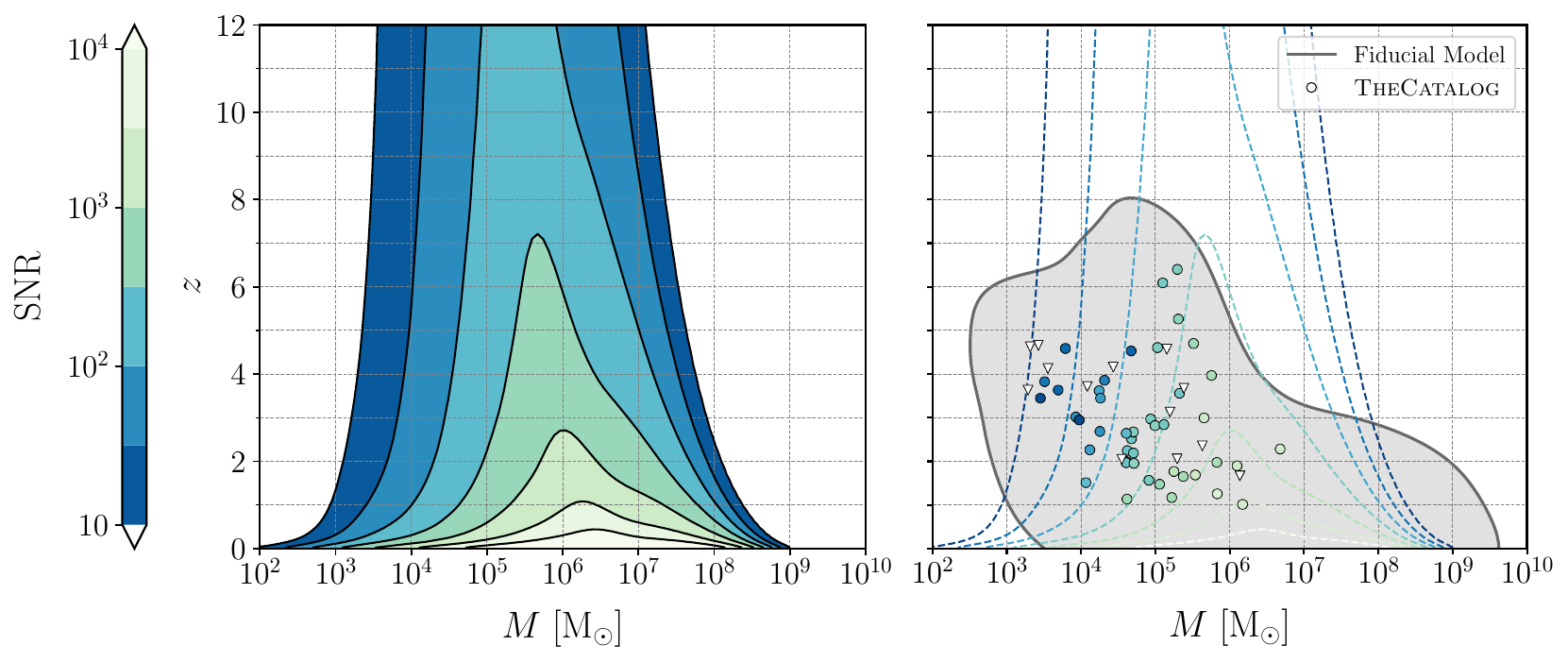}
    \caption{{\it Left panel.} Contour lines of constant SNR from massive BH binaries ($q=0.5$, $\chi_{1,2}=0.2$) detectable by LISA, adapted from Ref.~\cite{2024arXiv240207571C}. 
    The SNR shown on the color scale has been averaged over sky location, polarization and inclination assuming the dominant quadrupole emission mode. {\it Right panel.}  The gray contour encompasses 99\% of the estimated astrophysical BH population simulated with \textsc{L-Galaxies}. Circles indicate the 45 detectable sources of \textsc{TheCatalog}, colored according to their SNR [see step (vii) in Sec.~\ref{sec:lisacat}]. Of note, two pairs of sources overlap in this plot: one, located at $z\sim4.5$ with $M\sim 6\times 10^{3} {\rm M}_{\odot}$, have the same weights $w_{i}$ [i.e. identical source parameters, see step (iii) in Sec.~\ref{sec:lisacat}]; the other pair, located at $z\sim2$ with $M\sim 4\times 10^{4} {\rm M}_{\odot}$, are characterized by very similar parameter values. White triangles indicate sources with SNRs below the detection threshold. Note that undetectable sources with high masses ($M>10^4 \rm{M_{\odot}}$) are affected by the cut-off on the time to merger [step (ii) in Sec.~\ref{sec:lisacat}]. The dashed contour lines of constant SNR are the same of the left panel.}
    \label{fig:waterfallplot}
    \end{figure*}
The dashed distributions in all the three panels of Fig.~\ref{fig:catalogs_distribution} show the detectable events modeled using the additional modes ($l, |m|)=\{(2,1), (3,3), (3,2) (4,4)\}$ in the waveform approximant and removing the condition on $f_{\rm cut}$ [step (iv)].
We find higher-order modes marginally increase the number of detectable sources (see Ref.~\cite{2023MNRAS.525.2851S}). 
This is more significant for massive binaries with $M\gtrsim~10^{8}{\rm M_\odot}$, where the (2,2) mode frequency falls out of band. As shown in Fig.~\ref{fig:catalogs_distribution}, this preferentially impacts binaries hardened via stellar processes: including higher-order modes in LISA analyses might have important repercussions on the astrophysical interpretation of the data. 

In the following, we study in detail one specific realization of our simulated LISA catalogs, which we refer to as \textsc{TheCatalog}.
This specific catalog is close to the medians of the distributions in Fig.~\ref{fig:catalogs_distribution} and can thus be taken as representative. 
The 45 detectable sources of \textsc{TheCatalog} are shown as colored circles in the right panel of Fig.~\ref{fig:waterfallplot} together with the astrophysical population from which they are extracted. 
The left panel puts these sources into context by showing the common ``waterfall plot'', i.e. the averaged SNR in the mass-redshift space~\cite{2024arXiv240207571C}. LISA is expected to detect massive BH binary systems in the mass range of $M_{\rm tot}\sim10^4-10^7~{\rm M_\odot}$ and out to $z\sim7$, covering an SNR range from $\ssim10$ to $\ssim3000$.
The 13 sources marked with white triangles in Fig.~\ref{fig:waterfallplot} are undetectable by LISA.
Of these, 9 sources with $M>10^4~{\rm M_\odot}$ merge after the LISA mission nominal duration, thereby limiting their SNR growth, cf. step (ii) in Sec.~\ref{sec:lisacat}. 
Conversely, the 4 sources with $M<10^4~{\rm M_\odot}$ 
merge within the mission lifetime $T_{\rm LISA}$ but remain with $\rm{SNR}<10$ nonetheless.
These undetectable sources are excluded from the analysis presented in Sec.~\ref{sec:inf}. 

\section{\label{sec:inf} Statistical inference}

We compare a single  GW events against distributions of simulated sources using the Bayesian formalism spelled out in Ref.~\cite{2023MNRAS.525.3986M}; see also Refs.~\cite{2022hgwa.bookE..45V,2019MNRAS.486.1086M}. 
Specifically, we quantify the relative degree of consistency between individual detections from \textsc{TheCatalog} and the two sub-population models of binary hardening: gas (G) and stellar (S).

Given the posterior distribution $p(\theta|d, {\rm U})$ from each individual GW event obtained with some uninformative priors $p(\theta|{\rm U})$, we compute the Bayes factor between models G and S as follows:
    \begin{equation}
        \mathcal{B}_{{\rm G}/{\rm S}}=\frac{\mathcal{B}_{{\rm G}/{\rm U}}}{\mathcal{B}_{{\rm S}/{\rm U}}}=\frac{\displaystyle \int{p(\theta|d, {\rm U})\frac{p(\theta|{\rm G})}{p(\theta|{\rm U})}}\dd\theta}{\displaystyle \int{p(\theta|d, {\rm U})\frac{p(\theta|{\rm S})}{p(\theta|{\rm U})}\dd\theta}}\,.
        \label{eq:bayes_factor_matt}
    \end{equation}
The astrophysical probability densities $p(\theta|{\rm G})$ and $p(\theta|{\rm S})$ act as new informative priors on the targeted parameters which we reconstruct from the simulated sources using Gaussian kernel density estimates (KDEs)~\cite{2015scott.book.....B}.
    
Parameter estimation is performed through \textsc{Balrog} on $\theta~\!\!\!\!=~\!\!\!\{\theta_{\rm \alpha}, \theta_{\rm \zeta}\}$ where $\theta_{\rm \alpha}=~\!\!\!\!\{\mathcal{M}_c, \delta\mu, \chi_1, \chi_2, d_{\rm L}\}$ represent the quantities used to perform the astrophysical model selection [Eq.~(\ref{eq:astro_bayes_factor})] and $\theta_{\rm \zeta}=~\!\!\{t_m, \phi_0, \psi, \sin\beta,\lambda, \cos\iota\}$ are the additional extrinsic parameters characterizing the modeled signals.
In particular, we use the redshifted chirp mass $\mathcal{M}_c$ and the dimensionless mass difference $\delta\mu=~(m_1-m_2)/(m_1+m_2)$ as they are the two mass parameters that enter the Post-Newtonian (PN) evolution at the leading- and next-to-leading order, respectively. 
We run full Bayesian inference on simulated data using the nested sampling algorithm~\cite{10.1214/06-BA127} as implemented in \textsc{Nessai}~\cite{2021PhRvD.103j3006W}.
All the injections are in zero noise and we choose uniform priors $p(\theta|{\rm U})$ on each parameter over either its entire definition domain or a range that is sufficiently large to enclose the entire posterior.
In particular, we classify sources into three \rm{SNR} intervals: low ($10-100$), moderate ($100-1000$), and high ($>\!1000$).
For each of these, we estimate prior limits by calibrating against a fiducial source selected at the lower boundary of that range.
Therefore, the terms $p(\theta_{\rm \alpha}|{\rm U})$, $p(\theta_{\rm \zeta}|{\rm U})$, $p(\theta_{\rm \zeta}|{\rm G})$, $p(\theta_{\rm \zeta}|{\rm S})$ [see step (vi) of Sec.~\ref{sec:lisacat}] are constant, we can factor them out the integrals of Eq.~(\ref{eq:bayes_factor_matt}) and evaluate the KDEs exclusively on $\theta_{\rm \alpha}$.
This approach significantly improves the computational efficiency. Equation~(\ref{eq:bayes_factor_matt}) thus simplifies to
\begin{equation}
        \mathcal{B}_{G/S}=\frac{\displaystyle\int{p(\theta_{\rm \alpha}|d, {\rm U})p(\theta_{\rm \alpha}|{\rm G})\dd \theta_{\rm \alpha}}}{\displaystyle\int{p(\theta_{\rm \alpha}|d, {\rm U})p(\theta_{\rm \alpha}|{\rm S})\dd \theta_{\rm \alpha}}}\,.
        \label{eq:astro_bayes_factor}
\end{equation}
Note that our detection statistics is approximates as GW detectability depends on the data realization (for detailed discussions on this point see e.g. Refs.~\cite{2024ApJ...962..169E,2024PhRvD.109f3013M}). We verified that all of our posterior samples have $\rm{SNR}>~\!\!10$ and $q>~\!\!0.01$, which \aur{ensures}{ensure} consistency with the adopted detection criterion.
Furthermore, it is important to note that posterior samples constrained to a parameter space $\theta_{\rm \alpha}$ that is not supported by the gas [stellar] probability density, i.e. $p(\theta_{\rm \alpha}|{\rm G})=0$ [$p(\theta_{\rm \alpha}|{\rm S})=0$], result in a formally null [infinite] value for the Bayes factor in Eq.~(\ref{eq:astro_bayes_factor}). 
As a concrete example, see the discussion on the spin distributions in Sec.\ref{sec:results}.

Crucially, one needs to account for selection effects. The detectability-conditioned Bayes factor is given by~\cite{2023MNRAS.525.3986M}
    \begin{equation}
        \mathcal{D}_{{\rm G}/{\rm S}} = \frac{P(\mathrm{det}|{\rm S})}{P(\mathrm{det}|{\rm G})}\mathcal{B}_{{\rm G}/{\rm S}},
\label{eq:bayes_factor_det_weight}
    \end{equation}
    where $P(\text{det}|{\rm S})$  and $P(\text{det}|{\rm G})$ are the fraction of sources from the sub-populations S and G, respectively, that can be detected  \textit{a priori}  given adopted detection statistic ${\rm SNR}>10$ and $q>0.01$. 
    Our numerical implementation follows that described in Ref.~\cite{2023MNRAS.525.3986M}.
\begin{figure}
    \centering
    \includegraphics[width=1.\columnwidth, keepaspectratio]{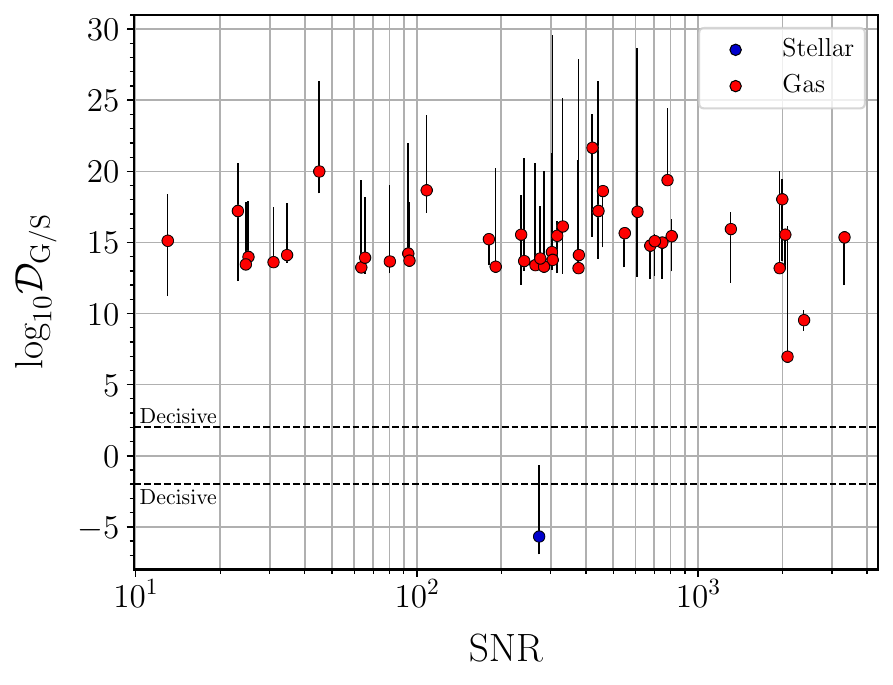}
    \caption{Logarithmic Bayes factor between the gas- and stellar- hardening hypotheses as a function of the source $\rm{SNR}$.
    Circles represent the 44 sources from \textsc{TheCatalog} used in the population analysis, with colors indicating their respective $f_{\rm{gas}}$ values.
    Horizontal dashed lines denote the threshold values of the Bayes factor according to the Jeffrey scale.
    Black error bars refer to the statistical (but not systematic) fluctuations associated to the underlying astrophysical model.
}
    \label{fig:bayes_factors}
    \end{figure}
    
\section{\label{sec:results} Results \& Discussion}

\subsection{Inference}

We apply the formalism from Sec.~\ref{sec:inf} to the 45 mergers in \textsc{TheCatalog}, which LISA is expected to detect.
Our parameter-estimation pipeline successfully recovers 44 sources, with posterior distributions on source parameters $\theta$ well confined within physically unbounded priors.
However, for one source with ${\rm SNR}\sim13$, the stochastic algorithm fails to converge, leading to poor-quality posteriors.
Therefore, we choose to exclude this source from the population analysis.
Additionally, sources with SNRs in the range~$10-100$ frequently exhibit non-linear correlations and multimodalities on $\theta_{\rm \zeta}$;
however both are sufficiently decoupled from the reconstructed $\theta_{\rm \alpha}$'s, which are the target parameters of our astrophysical model selection. In the following, we quote parameter estimates at 90\% confidence interval.    

Among the target parameters, $\mathcal{M}_c$ is the most precisely measured.
Specifically, we constrain it with a relative precision $\Delta\mathcal{M}_c/\mathcal{M}_c$ of $\ssim10^{-6}-10^{-4}$ for sources with moderate and high SNRs.
For low {\rm SNR}s, the relative precision is $~\ssim10^{-4}-10^{-3}$.
The spin components $\chi_{1}$ and $\chi_{2}$ are measured with a relative precision of $0.006\%-1\%$ at high and moderate \rm{SNR}s, and $1\%$ to $65\%$ at low \rm{SNR}s.

Six systems in the low$-~\!\!${\rm SNR} range exhibit mild biases on $\delta\mu$ due to the broad posteriors ($\Delta\delta\mu/\delta\mu\sim 6\%-160\%$ at 90\% credible level) and projection effects associated to non-linear parameter correlations.
Of these, three systems have posterior distributions that include the injected values only in their $99.9\%$ confidence inteval.
For the other three sources, the true injected value lies within the $98.5\%$ credible interval.
Finally, we emphasize that LISA will be able to measure the source luminosity distance with a relative precision $\Delta d_{\rm L}/d_{\rm L}$ of $\ssim0.3\%-2\%$ for high \rm{SNR}s, $\ssim1\%-45\%$ for moderate \rm{SNR}s, and $\ssim10\%-115\%$ for low {\rm SNR}s.

Uncertainties related to sky localization and host galaxy identification, which are crucial for multimessenger detections of gas-rich mergers~\cite{2022PhRvD.106j3017M}, will be addressed in future work.
 \begin{figure*}
    \centering
    \includegraphics[width=\textwidth, keepaspectratio]{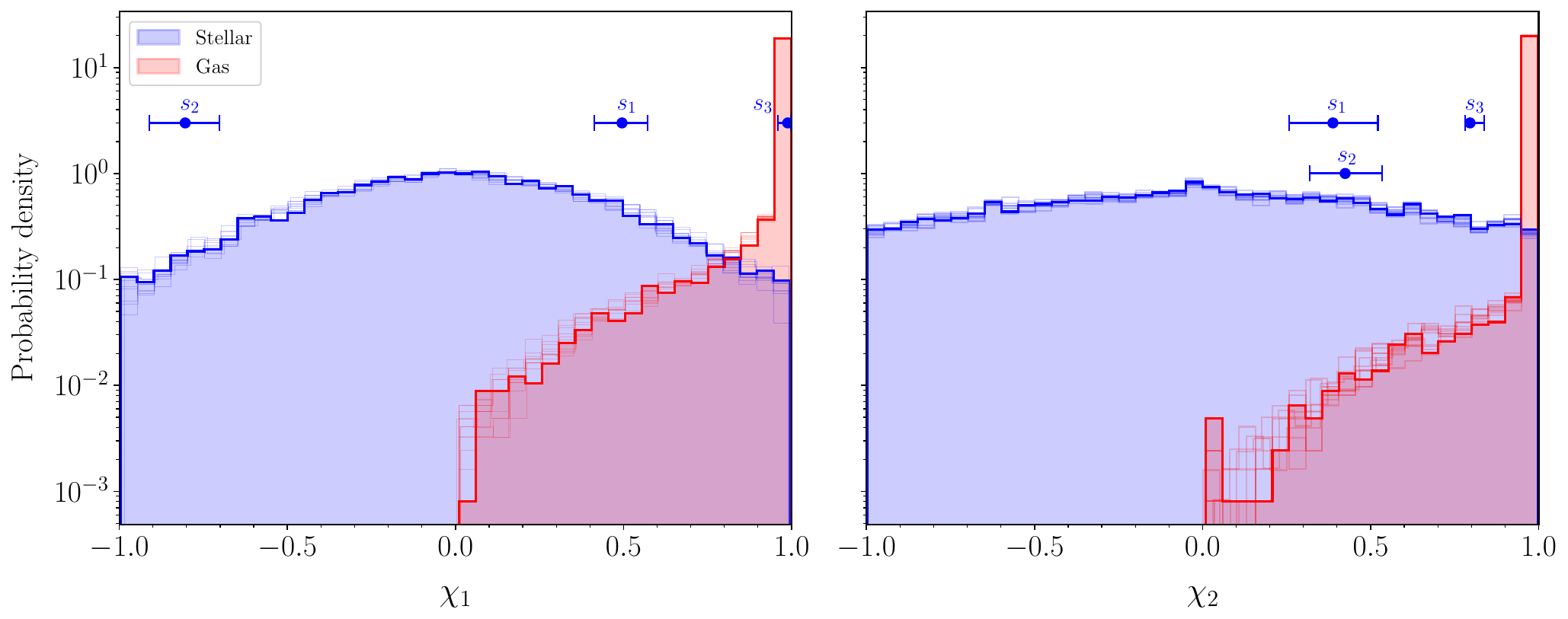}
    \caption{Marginal distribution of the aligned spin component of the primary BH  $\chi_{1}$ (left panel) and the secondary BH $\chi_{2}$ (right panel) for mergers in gaseous (red) and stellar (blue) environments.
    Thick histograms indicate our fiducial model; thin histograms indicate additional realizations.
    Blue dots denote the median value of source \textsc{$s_{1}$} from \textsc{TheCatalog} as well as the two additional sources \textsc{$s_{2,3}$}, c.f. Sec.~\ref{sec:results}.
    The error bars indicate the $90\%$ confidence intervals.
    The two sub-population distributions are jointly normalized to the full population.}
    \label{fig:spin12_fluctuations}
\end{figure*}

\subsection{Model selection}

We use the posteriors described above to evaluate the detection-conditioned Bayes factors ${\cal D}_{\rm{G/S}}$ of Eq.~(\ref{eq:bayes_factor_det_weight}).
Results are presented in Fig.~\ref{fig:bayes_factors}, where sources are ordered by their ${\rm SNR}$ and colored according to their true astrophysical sub-population.
For all the sources we find ``decisive'' evidence~\cite{24ce203a-855a-3aa9-952f-976d23b28943} in favor of the gas (stellar) sub-population model, with $\log_{\rm 10} \mathcal{D}_{\rm{G/S}}>2$ $(\log_{\rm 10} \mathcal{D}_{\rm{G/S}}<-2)$. 
Additionally, no significant correlation is observed between ${\cal D}_{\rm{G/S}}$ and the $\rm{SNR}$ of each source.
This is largely due to the highly precise measurement of the target parameters, as described above. 
Even the wider posterior distributions are well localized within the sub-population parameter space, so increasing the $\rm{SNR}$ does not lead to improved discrimination between the two astrophysical models.
We observe that most of the Bayes factors fall within the range $10\lesssim |\rm{log_{10}}\mathcal{D}_{\rm{G/S}}|\lesssim25$, with three exceptions that exhibit lower values.
Among theese, two sources with $\rm{SNR}\sim 2000$ and $7\lesssim |\rm{log_{10}\mathcal{D}_{\rm{G/S}}| \lesssim 10}$ evolved trough a gas-hardening phase ($f_{\rm{gas}}\sim 0.72$).
Conversely, the lowest Bayes factor ($|\rm{log_{10}}\mathcal{D}_{\rm{G/S}}|=5.63$) is associated with the only source in \textsc{TheCatalog} that hardened through stellar interactions ($f_{\rm gas}\sim 0.14$). 

We further test the analysis pipeline by generating ten additional realizations of our astropyhysical model to evaluate the impact of simulation uncertainties on the results.
The error bars in Fig.~\ref{fig:bayes_factors} indicate the resulting variability associated with each log-Bayes factor.
Even when accounting for these statistical fluctuations, model G is confidently favored for all the G sources, thus validating the previous results. 
For the one S source, model S remains favored, though to a lesser extent. 
Crucially, these error bars capture the statistical fluctuations in the underlying astrophysical model and not the systematic effects due to the many assumptions entering the model itself.
    
\subsection{Source parameters}

Based on these results, we now investigate how the astrophysical properties of massive BH binaries might influence the distinguishability between models G and S.
In Fig.~\ref{fig:popdensity_gas_st}, we show the 44 analyzed systems from \textsc{TheCatalog} together with the astrophysical sub-population distributions marginalized on the total mass and redshift. 
Notably, we observe that the gas-hardened sources are predominantly supported by the gas-hardening sub-population, with most falling within (outside) the $90\%$ contour level of the gas (stellar) distribution.
Conversely, the stellar-hardened source, hereafter referred to as \textsc{$s_{1}$}, falls outside the $90\%$ contour level of the stellar-hardening sub-population but is located close to the peak (i.e. within the $50\%$ contour level) of the gas-hardening sub-population.
This suggest that additional parameters may be impacting its $\rm{log_{10}}\mathcal{D}_{\rm{G/S}}$ value, favoring the stellar-hardening model.

To investigate further, in Fig.~\ref{fig:spin12_fluctuations} we analyze the one-dimensional distribution of the aligned component spins $\chi_{1}$ and $\chi_{2}$.
For the source \textsc{$s_{1}$} considered so far, the primary component spin $\chi_{1}=0.49^{+0.08}_{-0.08}$ lies below the $92^{\rm nd}$ and $1.5^{\rm th}$ percentiles of the stellar- and gas-hardening distributions, respectively, thus placing it confidently in the former.
Additionally, the secondary component spin $\chi_{2}=0.39^{+0.13}_{-0.13}$ is well below the $90^{\rm th}$ percentile of the stellar-hardening distribution and lies within only the  $0.3^{\rm rd}$ percentile of the gas-hardening distribution.
We find that the recovered $\delta \mu$ parameter does not provide significant support for model S, with the full posterior information falling below the $14^{\rm th}$ percentile of the stellar-hardening distribution but within the $50^{\rm th}$ percentile of the gas-hardening sub-population.

\subsection{Importance of the BH spins}

From the discussion above, we conclude that, at least for the astrophysical models considered here, the component spins are the crucial parameters for inferring the hardening processes of massive BH binaries.
We stress that the employed models describe a simplified scenario. 
Massive BH binaries evolving in gas-rich environments experience accretion torques that might align their component spins with the orbital angular momentum of the binary.
The models used here assume that the spin alignment process occurs rapidly, typically within the time-scale of the merger, thus leading to high and positive spin configuration at merger~\cite{2007ApJ...661L.147B} ($\chi_{1,2}\ssim0.98$ on average, as indicated by the red distributions in Fig.~\ref{fig:spin12_fluctuations}).
In contrast, binaries that evolve in gas-poor environments do not experience significant gas accretion and likely enter the GW dominated phase with roughly isotropic spin orientations, as reflected by the blue distributions in Fig.~\ref{fig:spin12_fluctuations}. 
Nevertheless, as pointed out in the literature, the overall picture is more complex due to the internal properties of the disk. 
Indeed, the effectiveness of disk alignment may be influenced by the gas temperature~\cite{2010MNRAS.402..682D}, accretion rate~\cite{2015MNRAS.451.3941G}, as well as the disk viscosity~\cite{2013MNRAS.429L..30L} and its initial angle of misalignment, which can lead to critical disk-breaking configurations~\cite{2013MNRAS.434.1946N}.
Accurate astrophysical models accounting for these properties are crucial for addressing such scenarios in future study.

Finally, we extend the analysis from \textsc{TheCatalog} by including two additional sources, \textsc{$s_{2}$} and \textsc{$s_{3}$}, selected from the LISA stellar-hardening distribution shown in Fig.~\ref{fig:catalogs_distribution}.
In order to provide a representative sample of LISA observations, we specifically select sources with \rm{SNR} in the range $100-1000$, which is most densely populated interval for detectable massive BH binaries in LISA (see Fig.~\ref{fig:waterfallplot}). 
This region is populated by sources with $M~\ssim 10^{4}-10^{6} \rm{M_{\odot}}$ at redshift $z\sim 2-4$.
Additionally, we select the two sources such that their primary spin components fall in the tails of the stellar-hardening sub-population from Fig.~\ref{fig:spin12_fluctuations}.
This allow us to further investigate the role of spins in the model comparison.

First, as shown in Fig.~\ref{fig:spin12_fluctuations}, the primary component spin $\chi_1=-0.8^{+0.1}_{-0.1}$ for source \textsc{$s_{2}$} is entirely constrained within a region where $p(\chi_{1}|\rm{G})=0$, thus making the KDE evaluation unnecessary, as pointed out in Sec.~\ref{sec:inf}.
Therefore, we can definitely favor model S over model G ($\rm{log_{10}\mathcal{D}_{G/S}}\!\!~\ll -2$) without even considering the influence of the other binary properties.
For source \textsc{$s_{3}$}, the analysis indicates a preference for the stellar-hardening scenario with $\rm{log_{10}\mathcal{D}_{G/S}}=-4.65$.
This case is particularly challenging because the gas-hardening model has also strongly support over the source posterior.
Specifically, \textsc{$s_{3}$} has a mass $M\sim 7\times 10^{4} M_{\odot}$ and is located at a redshift $z\sim 1.6$, positioning it at the peak of the $M-z$ distribution for the gas-hardening sub-population (see Fig.~\ref{fig:popdensity_gas_st}).
Additionally, it has a measured primary spin component of $\chi_1=0.98^{+0.01}_{-0.03}$.
This value is notably high and exceeds the $99.5^{\rm th}$ percentile of the stellar-hardening distribution and the $7^{\rm th}$ percentile of the gas-hardening distribution.
As illustrated in Fig.~\ref{fig:spin12_fluctuations}, \textsc{$s_{3}$} is situated in a high-density region for gas-hardened sources but in a relative sparse region for stellar-hardened sources.
Furthermore, the recovered value of $\delta\mu$ falls below the $9^{\rm th}$ ($38^{\rm th}$) percentile for the stellar (gas) sub-population, providing no further support for model S with respect model G.
On the other hand, the secondary component spin strongly favors model S.
In fact, the recovered value $\chi_2=0.79^{+0.04}_{-0.01}$ falls below the $95^{\rm th}$ ($1.2^{\rm nd}$) percentile for the stellar- (gas-) hardening sub-population.

This further emphasizes that, at least in the astrophysical setup considered here, component spins play a crucial role in constraining the hardening mechanisms responsible for the final coalescence of massive BH binaries, even in ambiguous but rare cases (less than $1\%$ of the stellar-hardening population sources detectable by LISA).
However, simulation-induced fluctuations reduce the log-Bayes factor of source \textsc{$s_{3}$} to as low as 0.35, indicating only a weak preference for model S.
Incorporating spin precession into the population analysis may be crucial to enhance model discrimination and gain further insights into the galactic environments where massive BH binary mergers occur.
This is left to future work.

\section{\label{sec:conclusions} Conclusions}   

In this work, we studied  the astrophysical origin of merging massive BH binaries detectable by LISA, with a specific focus on the hardening mechanisms that drive the binary toward the final coalescence.

Combining results from large-scale cosmological simulations and state-of-the-art models for massive BH binary evolution, we constructed a large set of realistic mock catalogs for LISA.
Our analysis showed that most detectable sources are expected to evolve in gas-rich environments.
We then performed Bayesian inference and model selection on a representative set of LISA detections and quantifies the statistical evidence in favor of either the gas-hardening or stellar-hardening channel.

Our results demonstrated that component spins might play a crucial role in distinguishing between these two astrophysical sub-populations and LISA will be able to provide clear insight given the expected precision on measured parameters.
This is a rather strong conclusion, which might however depend on the specific astrophysical model adopted in this paper. Reversing the argument, our findings imply that accurate modeling of the BH spins, both magnitudes and directions, is crucial to avoid major systematics in LISA's astrophysical inference.
    
Individual source inference with a richer waveform phenomenology --including precession effects, smaller mass-ratios, and potential eccentricity-- as well as more complete astrophysical simulations will  yield a more realistic estimate of the LISA potential to uncover the astrophysics of massive BHs.
For instance, in this paper we find that including higher-order modes in the adopted GW signal preferentially impacts sources that evolved through specific astrophysical formation pathways.
While we leave further explorations to future work, we stress that the methodology presented here is general and can be readily applied to more sophisticated models.

\begin{acknowledgments}

Data supporting this paper (including all mock LISA catalogs and  posterior samples) are publicly released at Ref.~\cite{datarelease}.
We thank Matthew Mould, Monica Colpi, Alberto Sesana, Christopher Moore, Nathan Steinle, and Jonathan Gair for  discussions.
A.S. and D.G. are supported by ERC Starting Grant No.~945155--GWmining, 
Cariplo Foundation Grant No.~2021-0555, MUR PRIN Grant No.~2022-Z9X4XS. 
A.S., R.B. D.I.-V., and D.G. are supported by MUR Grant ``Progetto Dipartimenti di Eccellenza 2023-2027'' (BiCoQ),
and the ICSC National Research Centre funded by NextGenerationEU.
R.B. is supported by Italian Space Agency Grant ``Phase A activities for the LISA mission'' %
No.~2017-29-H.0.
G.P. is supported by the Royal Society University Research Fellowship URF{\textbackslash}R1{\textbackslash}221500 and RF{\textbackslash}ERE{\textbackslash}221015.
D.G. is supported by MSCA Fellowships No.~101064542--StochRewind and  No.~101149270--ProtoBH.
Computational work was performed at University of Birmingham BlueBEAR High Performance Computing facility, at CINECA with allocations through EuroHPC Benchmark access call grant EUHPC-B03-24, INFN, and Bicocca, and at Google Cloud  through award No.~GCP19980904.
    
\textit{Software}:
We acknowledge usage of 
of the following 
\textsc{Python}~\cite{10.5555/1593511} 
packages for modeling, analysis, post-processing, and production of results throughout:
\textsc{Nessai}~\cite{2021PhRvD.103j3006W},
\textsc{matplotlib}~\cite{2007CSE.....9...90H},
\textsc{numpy}~\cite{2020Natur.585..357H},
\textsc{scipy}~\cite{2020NatMe..17..261V}.
\end{acknowledgments}
    
\bibliography{hardening}%
    
\vfill
\clearpage
\onecolumngrid
\newpage
\appendix
    
\end{document}